\documentclass[espf]{aastex}
\usepackage{emulateapj5}

\shorttitle{Mid-Infrared Emission from Hot Molecular Core Candidates}
\shortauthors{De Buizer et al.}

\begin{document}

\title{A Search for Mid-Infrared Emission from Hot Molecular Core Candidates }

\author{J. M. De Buizer\altaffilmark{1, 2, 3}}
\affil{Gemini Observatory, Casilla 603, La Serena, Chile}
\email{jdebuizer@gemini.edu}
\author{J. T. Radomski, C. M. Telesco\altaffilmark{2}, and R. K. Pi\~{n}a\altaffilmark{2, 4}}
\affil{Department of Astronomy, University of Florida, Gainesville, FL 32611, USA}

\altaffiltext{1}{Visiting Astronomer, Gemini North Observatory and NASA Infrared Telescope Facility.}
\altaffiltext{2}{Visiting Astronomer, W. M. Keck Observatory.}
\altaffiltext{3}{On leave from Cerro Tololo Inter-American Observatory, Casilla 603, La Serena, Chile}
\altaffiltext{4}{Present address: Photon Research Associates, Inc., 5720 Oberlin Drive, San Diego, CA 92121, USA}

\begin{abstract}
We present here mid-infrared images of seven sites of water maser emission thought to be associated with the hot molecular core (HMC) phase of massive star formation. Observations were obtained at the NASA InfraRed Telescope Facility 3-m, the Gemini 8-m, and Keck II 10-m telescopes. We have detected mid-infrared sources at the locations of two HMC candidates, G11.94-0.62 and G45.07-0.13. We observed G19.61-0.23 and G34.26+0.15, each of which have HMCs previously detected in the mid-infrared. We did not detect mid-infrared emission from either HMC source, and we place new upper limits on the mid-infrared flux densities for these HMCs that are much lower than their previously reported flux densities. We were able to obtain extremely accurate astrometry for our mid-infrared images of G9.62+0.19, and conclude that the mid-infrared emission thought to be coming from the HMC in this field is in fact coming from a different source altogether.           
\end{abstract}


\section{Introduction}

Hot molecular cores (HMCs) are believed to represent an extremely early stage of massive stellar birth. An HMC consists of a massive protostar surrounded by a thick envelope of accreting dust and gas. Though there are very few observations of these objects to date, we are beginning to piece together an empirically derived list of the basic observational characteristics of HMCs: 1) they are compact sources seen in radio-wavelength ammonia (or molecular line) images but have little or no radio continuum emission; 2) they lie in massive star forming regions near ultracompact H{\scriptsize II} (UC H{\scriptsize II}) regions; 3) they are too young and/or embedded to be seen in the optical or near infrared; and 4) they are often coincident with water maser emission. Only a small number of sources exist that have had such a holistic set of observations performed.

The work by Cesaroni et al. (1994) remains the most comprehensive work on the subject, even though their work involved radio-wavelength ammonia line imaging towards only four fields of massive star formation. In all four cases they found compact ($\sim$0.1 pc) ammonia clumps offset from, but near to, UC H{\scriptsize II} regions. In all four cases the water masers were directly coincident with the ammonia emission rather than the radio continuum emission of the UC H{\scriptsize II} region. Cesaroni et al. (1994) observed in the high-excitation (4,4) line of NH$_3$, which traces gas with kinetic temperatures of 50-200 K, and densities approximating n$_{H_2}$$\sim$ 10$^7$ cm$^{-3}$ (n$_{NH_3}$/n$_{H_2}$ $>$ 10$^{-6}$) within these HMCs.   

Cesaroni et al. (1994) also argue that gas and dust would be well mixed in the HMCs and  that there would be a high rate of collisions between the dust and gas. When densities in excess of n$_{H_2}$ $\gtrsim$ 10$^5$ cm$^{-3}$ exist, temperature equilibrium between gas and dust is established (Krugel and Walmsley 1984). Given the high observed densities of HMCs, it can be concluded that the gas kinetic temperature is a fair approximation to the dust temperature. The observed gas temperatures for HMCs is typically between 50 and 200 K (Cesaroni et al. 1994). Therefore one should be able to detect new HMCs via their mid-infrared emission, at least for the warmest sources. 

Recently there was a detection of mid-infrared emission from the HMC of G29.96-0.02 (De Buizer et al. 2002), a source that was first imaged by Cesaroni et al. (1994) in NH$_3$ emission. This source is also in the survey of Hofner \& Churchwell (1996), who studied the relationship between the 2 cm radio continuum emission of UC H{\scriptsize II} regions and the location of the water masers on the field. They found that G29.96-0.02, like several other sources in their survey, did not have water masers associated with the observed UC H{\scriptsize II} regions. Instead, the water masers tend to be in groups that lie offset from the radio continuum emission. The hypothesis is that the water masers are excited by embedded massive stellar sources within the HMCs. It is pointed out by Hofner \& Churchwell (1996) that several of the UC H{\scriptsize II} regions in their survey are cometary-shaped, like G29.96-0.02, and have their water masers located in front of the cometary arc of radio continuum emission.

Another source in common between the Cesaroni et al. (1994) survey and Hofner \& Churchwell (1996) survey is G9.62+0.19. Hofner \& Churchwell (1996) showed that both sources display radio continuum regions with water masers offset, and Cesaroni et al. (1994) showed that both have HMCs at the water maser locations as seen in NH$_3$.     

In order to extend our knowledge of the earliest stages of massive star formation, and understand more about the nature of HMCs, we have undertaken mid-infrared observations of several fields from the survey of Hofner \& Churchwell (1996) in an attempt to find more mid-infrared-bright HMCs. We concentrate on the fields that have UC H{\scriptsize II} regions with water masers well offset. Some of these fields are the sites of already detected HMCs at the water maser locations (e.g. imaged in molecular lines of NH$_3$, or already imaged in the mid-infrared). For the other fields, observations of molecules found in HMCs but not UC H{\scriptsize II} regions can be used to deduce the physical presence of HMCs. Fields that do not have directly detected HMCs were checked for spectroscopic observations of molecular species known to be tracers of the densest and warmest material. In particular, methyl cyanide (CH$_3$CN) and high excitation lines of NH$_3$ appear to be confined to those HMCs that are luminous at far-infrared and submillimeter wavelengths. From the list of Hofner \& Churchwell (1996) we have therefore isolated seven HMCs and/or HMC candidates for detailed imaging in the mid-infrared. While all sources were observed on the IRTF 3-m, others needed high spatial resolution or deep mid-infrared imaging as follow-up. These sources were observed on the Gemini 8-m and Keck II 10-m telescopes.  

\section{Observations}

Exploratory observations were first performed using the University of Florida mid-infrared camera and spectrometer, OSCIR, in 1997 September at the 3-meter NASA InfraRed Telescope Facility (IRTF). OSCIR employs a Boeing/Rockwell\footnote{Both Boeing and Rockwell are now subsidiaries of DRS Technologies.} 128 x 128 Si:As BIB (blocked impurity band) detector array, which is optimized for wavelength coverage between 8 and 25 \micron. The field of view of the array is 29$\arcsec$ $\times$ 29$\arcsec$, for a scale of 0.223$\arcsec$/pixel. Observations were centered on the H$_{2}$O maser coordinates given by Hofner \& Churchwell (1996), with 60 sec on-source exposure times taken through a broad-band \emph{N} filter ($\lambda _{0}$ = 10.46 $\micron$, $\Delta \lambda $ = 5.1 $\micron$) and the \emph{IHW18} (International Halley Watch, $\lambda _{0}$ = 18.06 $\micron$, $\Delta \lambda $ = 1.7 $\micron$) filter with a 30$\arcsec$ N-S chop throw. Unfortunately, cirrus clouds terminated the survey after only three sites were imaged.
 
The full survey was performed with deeper imaging in 2002 June, again at the IRTF, but this time using the Jet Propulsion Laboratory mid-infrared camera, MIRLIN. This instrument employs a Boeing\footnotemark[\value{footnote}] HF-16 128 x 128 Si:As BIB detector array. The pixel-scale is 0.475\arcsec/pixel, for a field of view of 61\arcsec\ $\times$ 61\arcsec. Observations were taken through the \emph{N4} ($\lambda_0$ = 11.70 $\micron$, $\Delta\lambda$ = 1.11 $\micron$) and \emph{Q3} ($\lambda_0$ = 20.81 $\micron$, $\Delta\lambda$ = 1.65 $\micron$) filters with on-source exposures times of 184 and 192 sec, respectively. All observations were taken at airmasses $<$1.5 under clear skies with low relative humidity ($<$25\%) with a chop throw of 61$\arcsec$ N-S. Sky and telescope subtraction were removed using the standard chop-nod technique, and the chop throw was large enough that there was no contamination from sources in the reference chop beam. The primary standard star used throughout the observations was $\gamma$ Aql, for which the flux densities were taken to be 61.1 Jy at 11 $\micron$ and 19.5 Jy at 20 $\micron$.      

Follow-up high-resolution observations were performed on two separate occasions. First, an observation of G9.62+0.19 was performed in 1999 April at the W.M. Keck II 10-m telescope. This observation was made using the OSCIR camera, a visiting instrument at the time. A single exposure was made of G9.62+0.19, with an on-source integration time of 240 sec through the \emph{IHW18} filter. At Keck, OSCIR has a field of view of 8$\arcsec$ $\times$ 8$\arcsec$, for a scale of 0.0616$\arcsec$/pixel. On the second occasion in 2001 May, observations were made at the Gemini North 8-m telescope of G19.61-0.23 and G34.26+0.15, again using OSCIR. One image was taken of G19.61-0.23 through the \emph{IHW18} filter for 180 sec. Images of G34.26+0.15 were taken through both the \emph{N} and \emph{IHW18} filters with 120 sec exposure times. At Gemini North, OSCIR has a field of view of 11$\arcsec$ $\times$ 11$\arcsec$, for a scale of 0.084$\arcsec$/pixel. Like the IRTF  observations, all of these high-resolution observations were taken at airmasses $<$1.5 under clear skies, and sky and telescope subtraction were removed using the standard chop-nod technique. Chop throws were 10$\arcsec$ N-S at Keck and 15$\arcsec$ E-W at Gemini. The standard star used at Keck and Gemini North was $\alpha$ Lyr, with an assumed flux density of 36.4 Jy at 10.5 $\micron$ and 12.0 Jy at 18.1 $\micron$. 

Table 1 lists the seven target fields in this survey. The table gives for each target field the coordinates of the water maser group number from Hofner \& Churchwell (1996) that is either associated with a known HMC, or is the likely location of the HMC candidate. The water maser positions of Hofner \& Churchwell (1996) were obtained with a connected element interferometer and have absolute positional uncertainties less than 0\farcs5. The mid-infrared absolute astrometry was obtained at IRTF by offsetting the telescope to the maser reference coordinates from nearby reference stars (typically less than 10$\arcmin$ away) whose coordinates were obtained from the Hipparcos Catalog. The absolute pointing uncertainty of the telescope was estimated to be better than 1\farcs5 in repeated trials, determined by offsetting between nearby Hipparcos reference stars. This astrometric accuracy is consistent with the pointing uncertainties previously achieved at IRTF during similar observations by Campbell et al. (2000).  We therefore conclude that the relative uncertainty of the astrometry between the maser reference features and our pointing is also better than 1\farcs5, in general. However, for each source detected we could confirm or increase this astrometric accuracy by comparing the mid-infrared field to the field imaged at other wavelengths. This process is discussed on a case-by-case basis for each source in \S3. 

The standard star images were used as point-spread function (PSF) stars for the IRTF observations. The measured resolution of the images was taken to be the average of the size of the full width at half maximum (FWHM) of the PSF stars imaged throughout the course of the observations. At IRTF the PSF observations yield an image resolution of 1\farcs2 at 10.5 $\micron$ and 1\farcs6 at 18.1 $\micron$ on the night using OSCIR, and 1\farcs3 at 11.7 $\micron$ and 1\farcs7 at 20.8 $\micron$ on two nights when MIRLIN was used. At Gemini and Keck, PSF stars were imaged near the location of each of the target fields imaged. The Gemini observations have a average measured resolution of 0\farcs48 at 10.5 $\micron$ and 0\farcs63 at 18.1 $\micron$, and PSF observations at Keck yielded a resolution of 0\farcs33 at 10.5 $\micron$ and 0\farcs41 at 18.1 $\micron$. 

Table 1 lists the observed flux densities of the HMC candidates at four wavelengths. The 10.5 and 18.1 $\micron$ flux densities correspond to the OSCIR data from various telescopes (which are annotated), and the 11.7 and 20.8 $\micron$ flux densities are from MIRLIN from IRTF. This paper concentrates discussion only on the HMC candidates on each field and reserves discussion of all other mid-infrared sources for future articles. However, since most of these fields have not been imaged in the mid-infrared before, Table 2 lists the flux densities for all mid-infrared sources found on all fields. Sources are designated by numbers which are incremented by increasing right ascension on the field\footnote{These numeric labels of the sources are shortened from the IAU recommended names which are of the form Glll.ll$\pm$b.bb:DRT03 \#, where 'l' and 'b' are the galactic latitude and longitude of the field and '\#' is the number of the source listed in Table 2. For instance, the first source of listed in Table 2 is G9.62+0.19:DRT03 1.}. The position of these sources are given as offsets in arcseconds from the position of the HMC on the field as given in Table 1. The flux densities listed in both tables are given with no quoted error. The largest sources of error in these values is the atmospheric variability the night of the observations and the uncertainty in the mid-infrared standard star flux density in each filter. The largest observed sky variability for any night of observations was approximately 10\% at 10.5 and 11.7 $\micron$ and 15\% at 18.1 and 20.8 $\micron$. These values were obtained from the variability of the measured detector counts for a given standard star throughout the night. We therefore assign the derived flux densities in this paper the above stated errors as conservative estimates of the actual photometric errors for all observations.  

\begin{table*}
\begin{minipage}{175mm}
\scriptsize
\caption{HMCs and HMC Candidates Observed}
\begin{tabular}{lccccccccc}
\hline
Site & Maser  & RA & Dec & RA & Dec & $F_{10.5\tiny{\micron}}$ & $F_{11.7\tiny{\micron}}$ & $F_{18.1\tiny{\micron}}$ & $F_{20.8\tiny{\micron}}$\\
& Group\tablenotemark{\dagger} & (B1950) & (B1950) & (J2000) & (J2000) & (mJy) & (mJy) & (mJy) & (mJy)\\ 
\hline 
G9.62+0.19   & 7 & 18 03 16.08 & $-$20 31 59.2 & 18 06 14.82 & $-$20 31 38.4 & $<$15\tablenotemark{c} & $<$25 & $<$48\tablenotemark{k} & \nodata\\
G11.94$-$0.62 & 3 & 18 11 03.70 & $-$18 54 18.4 & 18 14 00.27 & $-$18 53 23.6 & \nodata & 180 & \nodata & 829\\
G12.21$-$0.10  & 2 & 18 09 43.76 & $-$18 25 06.7 & 18 12 39.70 & $-$18 24 17.8 & \nodata & $<$28 & \nodata & $<$355\\
G19.61$-$0.23  & 6 & 18 24 50.05 & $-$11 58 31.6 & 18 27 37.84 & $-$11 56 37.1 & $<$10 & $<$31 & $<$17\tablenotemark{g} & $<$308\\
G34.26+0.15  & 2 & 18 50 46.28 & +01 11 12.5 & 18 53 18.62 & +01 14 58.1 & $<$4\tablenotemark{g} & $<$\em{?} & $<$21\tablenotemark{g} & $<$\em{?}\\
G45.07$-$0.13  & 2 & 19 11 00.41 & +10 45 43.3 & 19 13 22.01 & +10 50 53.8 & 4260\tablenotemark{\ddagger} & 5500\tablenotemark{\ddagger} & \nodata & 19700\tablenotemark{\ddagger}\\
G75.78+0.34  & 1 & 20 19 51.88 & +37 17 00.3 & 20 21 43.92 & +37 26 38.0 & \nodata & $<$26 & \nodata & $<$286\\
\hline
\end{tabular}
\vspace{-0.8cm}
\tablecomments{All 10.5 and 18.1 $\micron$ data were taken with OSCIR, and all 11.7 and 20.8 $\micron$ data were taken with MIRLIN. All data are from the IRTF unless otherwise noted. All values quoted with a ``$<$'' are upper limit flux densities at a 95\% confidence level. Flux density errors are taken to be 10\% at 10.5 and 11.7 $\micron$ and 15\% at 18.1 and 20.8 $\micron$ unless otherwise noted.}
\tablenotetext{\dagger}{Maser group designation from Hofner \& Churchwell (1996) that is at the HMC location, or the assumed maser location of the HMC candidate.}
\tablenotetext{c} {CTIO data from De Buizer, Pi\~{n}a, \& Telesco (2000).}
\tablenotetext{g} {Gemini data.} 
\tablenotetext{k} {Keck II data.}
\tablenotetext{?} {Even if the HMC was detected in the higher resolution Gemini images, the IRTF/MIRLIN images of this region would not be of sufficient resolution to resolve the HMC from source C.}
\tablenotetext{\ddagger} {There is some overlap of the extended emission from a nearby source with the HMC emission. The sources are resolved enough that a value can be estimated, however we estimate the error in the flux densities to be 5\% higher in each filter.}

\end{minipage}
\end{table*}       

\section{Results for Individual Sources}

\subsection{G9.62+0.19} 

G9.62+0.19 and its surrounding environment have been well studied at a variety of wavelengths (see Testi et al. 2000 and references therein). This complex region contains a wealth of high mass sources of different evolutionary states, from the HMC phase to well-developed H{\scriptsize II} regions. A study of the centimeter continuum emission from this region by Garay et al. (1993) yielded the designation of radio sources labelled A to E. A and B are large, extended ($\simeq30\arcsec$) regions of centimeter radio continuum emission, and D is a bright, compact radio continuum source just to the east of B. D is the southernmost component to a string of radio continuum sources that run to the northwest, ending with source C, approximately 20$\arcsec$ from D. The HMC, which lies in this string and is nearest to D, was first observed in thermal ammonia line emission by Cesaroni et al. (1994) and has been given the designation F. Masers of several species (H$_2$O, OH, CH$_3$OH, and NH$_3$) lie along this string of radio sources, and the HMC is coincident with several water masers.

This region was previously observed by De Buizer, Pi\~na, \& Telesco (2000) in the mid-infrared and a bright, compact source was found at both 10.5 and 18.1 $\micron$, as well as extended emission from H{\scriptsize II} region B. The bright mid-infrared source was given the designation G9.619+0.193:DPT00 1 (Figure 1). Lacking accurate astrometry and given the close proximity of G9.619+0.193:DPT00 1 with the extended emission to the east, De Buizer, Pi\~na, \& Telesco (2000) assumed that G9.619+0.193:DPT00 1 was mid-infrared emission from radio source D. Recently, Persi et al. (2003) claim that the astrometry of their mid-infrared images shows G9.619+0.193:DPT00 1 to be coincident with the location of the HMC. They established their astrometry by measuring the offset from the centroid of the mid-infrared emission from source B to the location of G9.619+0.193:DPT00 1. They then found the absolute position of G9.619+0.193:DPT00 1 by applying this measured offset to the published centroid of B as given in the MSX satellite catalog. They quote an optimistic uncertainty of 1-2$\arcsec$ using this method, and they state that the near-infrared emission they have observed is also coming from the HMC. A similar claim of near-infrared emission from the HMC was made previously by Testi et al. (1998). 

The MIRLIN observation of this region presented in Figures 1, 2, and 3 show something different than the previous mid-infrared observations of De Buizer, Pi\~na, \& Telesco (2000) or Persi et al. (2003). As can be seen in Figures 1 and 2, the MIRLIN observations at 11.7 $\micron$ revealed not only sources G9.619+0.193:DPT00 1 and B, but C and E as well. Figure 2 shows an overlay with the 3.5 cm radio continuum image of Phillips et al. (1998), which has a comparable field of view. Given the comparable resolutions and fields of view, as well as the fact that there are multiple sources in the field, positive identification can be made for all mid-infrared sources in the field. Figure 2 shows that the radio continuum and mid-infrared emission from the extended source B have similar peaks and extent. The radio continuum and mid-infrared peaks also coincide for sources C and E, as well as a source (here named G9.62+0.19:DRT03 9) which lies $\sim8\arcsec$ southeast of radio source D. However, one can see (Figures 2 and 3) that the position of G9.619+0.193:DPT00 1, given this extremely accurate registration, does not correspond spatially to \emph{either} the HMC location (F) or the peak of radio source D. Given the remarkably close coincidence of mid-infrared and radio continuum emission for sources B, C, E, and G9.62+0.19:DRT03 9, we claim the relative astrometry between the two wavelengths has a realistic uncertainty of $<1.0\arcsec$.

Figure 3 compares the higher resolution radio continuum and ammonia line emission observations from Hofner et al. (1994) with our mid-infrared image. One can see clearly that the mid-infrared emission of G9.619+0.193:DPT00 1 lies in between the compact radio continuum source D and the extended area of thermal ammonia emission F thought to encompass the HMC. Figure 4 shows the Keck 18.1 $\micron$ image of G9.619+0.193:DPT00 1 overlaid with the radio continuum of Testi et al. (2000) and the thermal ammonia emission of Hofner et al. (1994). The Keck image was registered with the peak location of G9.619+0.193:DPT00 1 in the IRTF image. It can be seen that there is no mid-infrared emission detected at Keck at the location of the HMC at a 3-$\sigma$ upper limit of 48 mJy at 18.1 $\micron$. 

Given the source crowding in this small region, it is most likely that the near infrared emission seen by Persi et al. (2003) is coming from the same source as the mid-infrared emission rather than the HMC location.

\subsection{G11.94-0.62}

This site is an analog to G29.96-0.02 in that it has a cometary shaped UC H{\scriptsize II} region present, with isolated water masers located out in front of the cometary arc (Hofner \& Churchwell 1996). There are two clusters of water masers near the UC H{\scriptsize II} region, and both are offset to the west. Maser group 2 lies $\sim$2$\arcsec$ from the radio continuum peak of the UC H{\scriptsize II} region, and therefore it was expected that it would be difficult to image emission of a hot molecular core separate from the emission of the UC H{\scriptsize II} in the mid-infrared, if one existed there. Water maser group 3, however, is located $\sim10\arcsec$ from the UC H{\scriptsize II} region peak, and therefore represents a good hot molecular core candidate that can be searched for in the mid-infrared.       

Figures 5 and 6 show the MIRLIN observations of this site at 11.7 and 20.8 $\micron$, respectively. Interestingly, the cometary shape seen in the cm radio continuum images of Hofner \& Churchwell (1996) is not evident in the mid-infrared (Figure 7). This is also the case for the 2.7 mm images of Watt \& Mundy (1999). The astrometry for this particular image is fairly well determined, since pointing of the telescope was checked immediately before and after the observations and was found to be reproducible to better than 1$\arcsec$. Therefore, the overlay of the image by Hofner \& Churchwell (1996) in Figure 7 is considered to be at least that accurate. The goodness of the astrometry is further indicated by several matching structures in both the mid-infrared and radio continuum, however the radio peak is surprisingly not present in the mid-infrared. 

As can be seen in Figure 7, we detected a mid-infrared source at the location of water maser group 3. Flux densities for this source at 11.7 and 20.8 $\micron$ are given in Table 1. The mid-infrared emission from this candidate HMC is compact and unresolved, and exactly coincident with the water masers. 

G11.94-0.62 has not yet been imaged in molecular line emission (e.g. ammonia) with sub-arcsecond resolution. The $\sim3\arcsec$-resolution 2.7 mm images by Watt \& Mundy (1999) of this site do not show any dust continuum emission towards the location of maser group 3. They also failed to detect molecular CH$_3$CN or $^{13}$CO emission in their imaging of this site, even though both were detected spectroscopically by Churchwell, Walmsley, \& Wood (1992) along with CS emission. Watt \& Mundy (1999) speculate that, because of the array configuration of their interferometric observations, they may not have detected the molecular emission found by Churchwell, Walmsley, \& Wood (1992) if it is extended over a large emitting region. Cesaroni, Walmsley, \& Churchwell (1992) also detected NH$_3$ line emission towards G11.94-0.62 spectroscopically, but it cannot be ascertained whether the NH$_3$ is coming from an HMC at the location of water maser group 3, or the UC H{\scriptsize II} region, or both. It is encouraging that the molecular tracers, especially those from hot and dense gas (e.g. high-excitation NH$_3$) are present in the region, implying the presence of an HMC on the field. However, it is disconcerting that Watt \& Mundy (1999) do not see the HMC candidate in their molecular line imaging of CH$_3$CN, which is thought to be an excellent tracer of HMCs. 

\subsection{G12.21-0.10}

This is the site of another cometary UC H{\scriptsize II} region with several water maser groups out in front of the cometary arc. This region has been observed to contain several tracers of hot and dense chemically enriched gas, such as high-excitation NH$_3$ (Cesaroni, Walmsley, \& Churchwell 1992; Anglada et al. 1996), CH$_3$CN (Millar \& Hatchell 1997), and CS (Plume, Jaffe, \& Evans 1992), thus making it a prime HMC candidate. However, there are not yet any direct molecular line images of this site.

Observations of G12.21-0.10 using MIRLIN did not reveal the presence of mid-infrared emission from an HMC at the location of the water masers at either 11.7 or 20.8 $\micron$. Even more surprising is that the UC H{\scriptsize II} region, detected at cm radio wavelengths (Hofner \& Churchwell 1996) and in the submillimeter (Hatchell et al. 2000), was not detected at either mid-infrared wavelength using MIRLIN.

\subsection{G19.61-0.23}

The molecular and ionized environment of G19.61-0.23 was extensively studied by Garay et al. (1998). This site was found to contain five distinct H{\scriptsize II} regions seen in radio continuum emission (labelled A, B, CD, E and F). Additionally, Garay et al. (1998) found three ammonia clumps in the region. Water and OH masers were discovered to be associated with the middle and most prominent of the three ammonia clumps. 

In the work by De Buizer (2000, 2003), OSCIR/IRTF observations revealed the presence of mid-infrared sources coincident with all of the H{\scriptsize II} regions of Garay et al. (1998) except source F. Figure 8 shows the 18.1 $\micron$ observations of De Buizer (2000) overlaid with the radio continuum of Hofner \& Churchwell (1996). The registration between these overlaid images was performed by matching the inner radio continuum contours of Garay et al. (1998) with those of Hofner \& Churchwell (1996), which match well. Then the outer radio continuum contours of Garay et al. (1998), which are well matched with the IRTF mid-infrared contours, were used to register the mid-infrared image to the Hofner \& Churchwell (1996) image. The maser positions are known to be accurate to 0$\farcs$1 with respect to the radio continuum. The fact that there are four mid-infrared sources in common with the radio continuum sources of Garay et al. (1998) suggests that the astrometry is indeed accurate. Therefore the relative astrometric accuracy between the mid-infrared and the radio continuum images is believed to be $\lesssim1\farcs5$. 

In Figure 8 one can see that there is a weak, mid-infrared source coincident with the location of water maser group 6 from Hofner \& Churchwell (1996). This source was not detected at 10.5 $\micron$, and was a 3-$\sigma$ detection with a quoted flux density of 230$\pm$30 mJy at 18.1 $\micron$ (De Buizer 2000). However, this mid-infrared emission was considered to be real given the coincidence of the source with the water masers and the fact that this group of water masers exists within the confines of the middle ammonia clump of Garay et al. (1998) (Figure 9). Based on this coincidence of ammonia and water masers, De Buizer (2000) argued that this source is indeed an HMC.

Follow-up observations were performed at Gemini North using OSCIR and did not confirm the presence of a source at the location of water maser group 6 (Figure 9). A 3-$\sigma$ upper limit on the 18.1 $\micron$ flux density of a source in this area from these new observations is 11 times less than the quoted flux density for this source from De Buizer (2000). 

\subsection{G34.26+0.15}

G34.26+0.15 is a star forming complex that has been the subject of a wealth of observations at several wavelengths. It is outside the scope of this paper to discuss all of the aspects of how the new mid-infrared observations presented here complement the previous observations for all sources in this region. Such discussion will be saved for a future paper. We will, however, discuss here observations that pertain to the HMC, and we will briefly touch on certain issues regarding the other sources in the field.

Radio continuum observations by Hofner \& Churchwell (1996) show this to be a site of another cometary-shaped UC H{\scriptsize II} region with water masers out in front of the cometary arc. These observations show the same radio continuum sources found by Gaume, Fey, \& Claussen (1994), but with less extended emission than they see. Gaume, Fey, \& Claussen (1994) named the cometary UC H{\scriptsize II} region source C, and the two unresolved bright UC H{\scriptsize II} regions located east and in front of the cometary arc as A and B (southern and northern sources, respectively; see Figure 10). In front of the cometary arc of C, there is a confirmed HMC that is seen in several molecular species (e.g. Keto et al. 1992; Churchwell, Walmsley, \& Wood 1992). 

The most complete mid-infrared observations of this region were performed by Campbell et al. (2000) at the IRTF. They observed the region through six mid-infrared filters from 8.0 to 20.6 $\micron$ with 1$\farcs$2-1$\farcs$7 resolution. Like the earlier mid-infrared observations of Keto et al. (1992) of this site, Campbell et al. (2000) found the mid-infrared emission of radio continuum source C not to be cometary in the mid-infrared, and failed to detect emission from radio continuum source B. However, Campbell et al. (2000) did discover two new sources seen only at 20.6 $\micron$, one of which, designated E, lies just south of source C. Keto et al. (1992) also claimed to have discovered an additional mid-infrared source at 12.5 $\micron$ at the location of the HMC with a flux density of 70 mJy. Campbell et al. (2000) did not detect this source, but they could not discount its existence given the noise level of their images.

One problem with comparing the radio continuum images of G34.26+0.15 with our images is that the separation between sources A and C is 0$\farcs$5 larger in the radio than in our mid-infrared images. Campbell et al. (2000) notes this as well, stating that registering their mid-infrared images on radio source A resulted in an offset of 0$\farcs$5 in the peak locations of C that did not appear to be real. They concluded that the lower signal-to-noise ratio of A was likely to be the cause of the scatter, and they therefore registered all of their mid-infrared images using the radio continuum peak of C. In our mid-infrared images source A has good signal-to-noise, yet we have drawn the same conclusion that the mid-infrared morphology of source C is too similar to the radio morphology for the offset to be real. Therefore, like Campbell et al. (2000), we registered our mid-infrared images using source C. Surprisingly, the 12.5 $\micron$ image of Keto et al. (1992) shows sources A and C to have the same separation as the radio continuum sources, inconsistent with the conclusion by Campbell et al. (2000) and our results here. This uncertainty in registration, however, does not affect our following interpretation of the region.

Figures 11 and 12 show the 10.5 and 18.1 $\micron$ OSCIR images from Gemini North. Two things are immediately obvious from these figures. First, radio continuum source C does not look like a single stellar source with a cometary shaped UC H{\scriptsize II} region in the mid-infrared. Instead it consists of perhaps four individual components (not two, C and E, as suggested by Campbell et al. 2000; see Table 2). Second, the mid-infrared emission from the HMC is absent in both images. This can best be seen in Figure 13 which shows the Gemini 10.5 $\micron$ image overlaid with the 12.5 $\micron$ image of Keto et al. (1992). Figure 14 shows the NH$_3$ image of Keto et al. (1992) overlaid on the 18.1 $\micron$ image from Gemini for comparison. The NH$_3$ peak is actually a little south of the 12.5 $\micron$ peak, however the Gemini observations show that there are no signs of mid-infrared emission from either location or from the location of any of the other isolated water maser groups in the region. These Gemini observations cast doubt on the existence of the mid-infrared source seen by Keto et al. (1992), since the 3-$\sigma$ upper limit on the presence of such a source is 4 mJy at 10.5 $\micron$ and 21 mJy at 18.1 $\micron$. 

Keto et al. (1992) also presented one other mid-infrared detection of an HMC at the location of W3(H$_2$O). Similar to the result presented here, this other HMC was not detected in the recent observations by Stecklum et al. (2002), who give an upper limit far below the flux density quoted by Keto et al. (1992).         

\subsection{G45.07-0.13}

The observations of Hofner \& Churchwell (1996) show this site to contain an unresolved, bright UC H{\scriptsize II} region coincident with several water masers. There is, however, one tight group of water masers offset $\sim2\arcsec$ north of the UC H{\scriptsize II} peak. Given the compact nature of the UC H{\scriptsize II} region in the radio, the northern group of water masers was believed to be far enough away to distinguish mid-infrared emission from an HMC, if one exists there.

There exist no high-spatial resolution molecular line images of this site. However, like G11.94-0.62, this region has been observed to contain tracers of hot and dense chemically enriched gas, such as CH$_3$CN (Pankonin et al. 2001). The mid-infrared images presented in Figures 15 and 16 are 11.7 and 20.8 $\micron$ images, respectively, from MIRLIN at then IRTF. These images show three mid-infrared sources in the area. The middle and brightest of these mid-infrared sources is closest to the location of the UC H{\scriptsize II} region and assumed to be coincident. Figure 17 shows the 11.7 $\micron$ image overlaid with the radio continuum contours and masers from Hofner \& Chuchwell (1996). In this figure one can see that the location of the isolated northern water maser group is exactly coincident with the northern mid-infrared source.  

G45.07-0.13 has also been imaged at 3 mm by Hunter, Phillips, \& Menten (1997). They see emission from the UC H{\scriptsize II} region as well as the southern mid-infrared source. Given the resolution of the 3 mm images, it is difficult to say whether or not they should be able to resolve the emission from the HMC candidate at the northern water maser location. The HMC candidate may be faint at 3 mm, or unresolved from the UC H{\scriptsize II} region, or both.

The water masers associated with the UC H{\scriptsize II} region, as discussed in Hofner \& Churchwell (1996), lie in a linear distribution at position angle of -45$\arcdeg$. This is a similar position angle to the axis of an outflow seen at this location by Hunter, Phillips, \& Menten (1997) observed in CO and CS. Hofner and Churchwell (1996) suggest that the water masers and outflow are related, thus implying that the outflows here emanate from the UC H{\scriptsize II} region. Hunter, Phillips, \& Menten (1997) also claim that the outflow is driven by a single isolated stellar source at the center of the UC H{\scriptsize II} region. However, given the resolution of the CO and CS images, any of the three sources,including the HMC candidate, seen in the mid-infrared are candidates for the outflow source.

\subsection{G75.78+0.34}

The UC H{\scriptsize II} region present here is coincident with the OH maser ON2-N, and therefore this site is also known by that name. Furthermore, this site is also known as G75.78-0.34, as a result of a misprint in the paper by Hofner \& Churchwell (1996). At radio continuum wavelengths, Hofner \& Churchwell (1996) find the UC H{\scriptsize II} emission to be cometary shaped, again with water masers offset and located in front of the cometary arc by $\sim2\arcsec$. 

G75.78+0.34 has never been imaged in molecular line emission, however it is the site of a variety of molecules known to be tracers of high density material associated with HMCs. Tracers like CH$_3$CN (Hatchell et al. 1998) and CS (Olmi \& Cesaroni 1999) have been detected here, as well as high excitation lines of NH$_3$ (Olmi, Cesaroni, \& Walmsley 1993). However, the best evidence for the presence of a compact object at the location of the water masers is the detection of a 7 mm source at the maser location by Carral et al. (1997). Emission was also detected by Carral et al. (1997) at wavelengths out to 2 cm. Given the extremely high emission measure of this object, it has been deemed a ``hypercompact H{\scriptsize II} region'' (Kurtz \& Franco 2002), and as such is thought to represent a stellar phase much younger than the UC H{\scriptsize II} phase.

Despite the fact that this appears to be an excellent candidate for mid-infrared study, we failed to detect any mid-infrared emission at either 11.7 or 20.8 $\micron$ with MIRLIN at IRTF from the location of the water masers. Even more surprisingly, like G12.21-0.10, the UC H{\scriptsize II} region was not detected either. Upper limits for flux densities for these objects are given in Table 1.

\section{Physical Properties of the HMC Candidates}

We have derived estimates of several physical parameters for the newly detected HMC candidates seen in G11.94-0.62 and G45.07-0.13. A detailed discussion of how the color temperature, mid-infrared luminosity, and spectral type were derived are discussed in De Buizer, Pi{\~ n}a, \& Telesco (2000), and the discussion of the limitations of this method are discussed in more detail in De Buizer et al. (2002b). Basically, for each source a blackbody function was fit to the 11.7 and 20.8 $\micron$ flux densities. From this Planck-function fit a dust color temperature was found. Integrating this Planck function from 1 to 600 $\micron$ and using this distance to the source we derived a mid-infrared luminosity. The distances to the sources were taken to be 4.2 kpc and 6.0 kpc for G11.94-0.62 and G45.07-0.13, respectively (Hofner \& Churchwell 1996).

These calculations yield a blackbody dust color temperature of 167 K for G11.94-0.62 and 181 K for G45.07-0.13. Because these are blackbody dust color temperature values, they can be considered upper limits on the actual mid-infrared dust color temperature. The blackbody mid-infrared luminosities derived for the sources are 90 and 4360 L$_{sun}$ for G11.94-0.62 and G45.07-0.13, respectively. Again, because these are blackbody mid-infrared luminosities, they are a reflection of only a part of the bolometric luminosity and therefore can be considered to be lower limits (possibly extreme lower limits) to the actual bolometric luminosity of the central heating sources. Making the simple assumption that the mid-infrared luminosities are equal to the bolometric luminosities of the sources, we can derive ZAMS spectral type limits which would be an estimate of the latest spectral type the central heating source could possibly be. These spectral type limits were found to be B9 and B0 for G11.94-0.62 and G45.07-0.13, respectively.   

We can also derive limits to the physical size of the HMC candidates in the mid-infrared. The blackbody lower limit size calculated for G11.94-0.62 is 105 AU and 624 AU for G45.07-0.13. Following the procedure outlined in Soifer et al. (2000), we used seven IRTF/MIRLIN PSF observations to derive an upper limit size for the HMC candidates. The rms scatter of the PSF FWHMs was found to be 0$\farcs$1 at 11.7 $\micron$. Since the HMCs, like the PSF stars, are unresolved at this wavelength, we can take the value of 5-$\sigma$ (0$\farcs$5) as the upper limit of the size of both HMC candidates under the assumption that a difference of 5 times the standard deviation of the PSF FWHM would have been easily discernable. This yields a upper limit physical size for the HMC candidates of 2100 AU for G11.94-0.62 and 3000 AU for G45.07-0.13.

We can compare these to the only other mid-infrared bright HMC presently, G29.96-0.02. This source, however, is resolved with a PSF-quadrature-subtracted physical size of 6720 AU. This is a few times larger than the upper limits we have set on G11.94-0.62 and G45.07-0.13. Since the HMC in G29.96-0.02 is resolved, the dust color temperature will be closer to the quoted optically thin dust color temperature value of 105 K (De Buizer et al. 2002a) which yields a mid-infrared luminosity of 490 L$_{\sun}$. This estimates the spectral type at $>$B6, derived from the mid-infrared luminosity alone. This shows that all three HMCs are at least luminous enough to contain B-type stars, and very possibly O-type stars at there centers. This is consistent with the hypothesis that the central stars heating these mid-infrared bright HMC candidates are indeed massive, assuming that they are heated by a single source. 

\section{Discussion}

The main result of these observations is the detection of two new HMC candidates, which when combined with the mid-infrared detection of G29.96-0.02 (De Buizer et al. 2002a), represent three HMC candidates to be directly imaged recently in the mid-infrared from the survey of Hofner \& Churchwell (1996). Unlike G29.96-0.02, however, the HMC candidates of G11.94-0.62 and G45.07-0.13 have not been directly imaged in molecular line emission. Therefore can we be sure that these are, in fact, HMCs? There are several reasons why it is believed that these two mid-infrared sources are HMCs. First, all of the mid-infrared sources have no radio continuum emission, implying that they are indeed young and embedded. Second, this survey has relatively good astrometry and both of these HMC candidates are directly coincident with clusters of water masers, which are generally believed to trace young massive stars. Third, there are spectroscopic observations providing evidence of high-excitation molecular emission line coming from each of these regions, a necessary (but not sufficient) condition if these are HMCs. Finally, we derive estimates of mid-infrared luminosities for these HMCs that are consistent with their being heated by massive central stars. Taken together, these four pieces of evidence combined lead to the belief that these are indeed HMCs. However, confirmation of the hot molecular nature of these sources will have to come from high resolution molecular line imaging (e.g. NH$_3$ or CH$_3$CN).

While the mid-infrared detections of G11.94-0.62, G29.96-0.02, and G45.07-0.13 demonstrate that HMCs can indeed be imaged at mid-infrared wavelengths, our lack of confirmation of the previously reported detections of HMCs in the mid-infrared in the fields of G9.62+0.19, G19.61-0.23, and G34.26+0.15 show that one must be extremely careful in their astrometry and that one must image these sources deeply in order to confirm their existence as HMCs. Given the gregarious nature of young massive stars, the fields containing HMCs are crowded with sources at many different stages of formation. We must be careful that we are referring to the same source when comparing fields at different wavelengths, since these different wavelengths are sensitive to different stellar phases.   

Of the 21 sites in the survey of Hofner \& Churchwell (1996), nine stand out as good candidates for HMCs, given the fact that they contain water masers isolated and offset from nearby UC H{\scriptsize II} regions and display evidence for molecular line emission. G43.89-0.38 was the only one of these nine sources not observed in either this work or the G29.96-0.02 article by De Buizer et al. (2002a). Of the eight sources observed from Hofner \& Churchwell (1996), we have detected mid-infrared emission from the HMCs of three sources in total. Though it is certainly small number statistics, this suggests a mid-infrared detection success rate towards isolated water masers in regions of massive star formation of 38\%. Why is there such a low mid-infrared detection rate for HMCs? The most obvious reason is that if the HMCs are too cold and embedded, very little mid-infrared emission would be expected. This would be the case if the HMCs are at an extremely early stage of development. This implies that HMCs that display mid-infrared emission are at a more evolved stage of formation than HMCs lacking mid-infrared emission. 

Since we need to confirm the validity of the HMC through molecular line imaging anyway, one might ask why we should even image the HMCs in the mid-infrared at all. In addition to the reason just mentioned, i.e. that it gives us an idea of the relative age of the HMC, one may be able to use mid-infrared observations at more wavelengths to more accurately describe the physical characteristics of massive star formation. Models of the earliest stages of massive star formation are being produced which calculate emergent thermal dust spectral energy distributions (e.g. Osorio, Lizano, \& D'Alessio 1999). These SEDs model an HMC as an envelope of gas and dust that is infalling and accreting onto a massive central protostar. The models take into account the effects of mass accretion rate and the luminosity of the central proto-stellar object on the radiative transfer through the envelope as a function of the envelope size. 

The simple models of Osorio, Lizano, \& D'Alessio (1999) showed a great dependence of the emergent SED at mid-infrared wavelengths on the physical characteristics of the HMC. On the steep Wien side of the SED, the depth and shape of the 10 $\micron$ silicate feature and the lower wavelength fall-off of emission are heavily dependent on accretion rate and central proto-stellar luminosity. In the Rayleigh-Jeans portion of the SED (i.e. at far-infrared and longer wavelengths), the shape of the SED is simply a constant slope that appears to be predominantly determined by the dust opacity at these wavelengths. Most models adopt a power law for the dust opacity of the form, $\kappa _{\lambda}\propto \lambda ^{-\beta }$, where $1\leq \beta \leq 2$ for $\lambda \geq 200$ $\micron$. Consequently, data at wavelengths longer than $\sim$100 $\micron$ do not drastically effect the SED and hence do not reveal much about the physical processes, such as accretion rate. However, the molecular lines in the sub-millimeter and millimeter are still helpful for deriving the HMC temperatures (which tells one where the SED peaks) and density values needed for accurate modelling of the radiative transfer. Though such models are still in their infancy, a more complete sampling of the SED at a variety of mid-infrared wavelengths (i.e. more than the one or two wavelengths sampled here) will yield more accurate information of the physical characteristics of HMCs as the models become more refined.          

\section{Conclusions}

In this paper we presented the detection of mid-infrared emission from the locations of two HMC candidates: G11.94-0.62 and G45.07-0.13. Both of these sources are believed to be genuine HMCs because they display no radio continuum emission, are directly coincident with clusters of water masers, reside in regions of high-excitation molecular emission line emission observed spectroscopically, and have derived mid-infrared luminosities that imply they are being heated internally by massive B or O type stars.

We also presented deep mid-infrared images of two HMCs previously detected in the mid-infrared: G19.61-0.23 and G34.26+0.15. For both of these sources we failed to detect mid-infrared emission from the HMC location at any wavelength. For G9.62+0.19, we were able to determine that the mid-infrared emission thought to be associated with the HMC is in fact offset $\sim$2$\arcsec$ from the HMC location. Given the density of sources on the field, the emission seen in molecular line images and the mid-infrared emission are most likely coming from different sources altogether. 

The two HMCs newly detected in the mid-infrared, along with the HMC in G29.96-0.02 (De Buizer et al. 2002a), represent a small, but growing group of sources that can be modelled to learn about the physical characteristics of massive star formation. Further observations and mid-infrared spectra of these sources will provide extremely useful information for the modelling and characterization of these youngest stages of massive stellar birth.     

\acknowledgments 
Data presented in this article were in part obtained at the Infrared Telescope Facility, which is operated by the University of Hawaii under a cooperative agreement with the National Aeronautics and Space Administration. 
This material is based upon work supported by the National Aeronautics and Space Administration under Cooperative Agreement no. NCC 5-538 issued through the Office of Space Science, Planetary Astronomy Program.

Some data were also obtained at the Gemini Observatory, which is operated by the
Association of Universities for Research in Astronomy, Inc., under a cooperative agreement
with the NSF on behalf of the Gemini partnership: the National Science Foundation (United
States), the Particle Physics and Astronomy Research Council (United Kingdom), the
National Research Council (Canada), CONICYT (Chile), the Australian Research Council
(Australia), CNPq (Brazil) and CONICET (Argentina).

Further data presented herein were obtained at the W.M. Keck Observatory, which is operated as a scientific
partnership among the California Institute of Technology, the University of
California, and the National Aeronautics and Space Administration. The
Observatory was made possible by the generous support of the W.M. Keck
Foundation.

\clearpage

\begin{table*}
\begin{center}
\begin{minipage}{120mm}
\caption{MIR Flux Densities for All Sources Observed in Target Field of View}
\begin{tabular}{lccccccc}
\hline
Field & Source\tablenotemark{\dagger} & $\Delta$RA\tablenotemark{\dagger} & $\Delta$Dec\tablenotemark{\dagger} & $F_{10.5\tiny{\micron}}$ & $F_{11.7\tiny{\micron}}$ & $F_{18.1\tiny{\micron}}$ & $F_{20.8\tiny{\micron}}$\\
&  & ($\arcsec$) & ($\arcsec$) & (Jy) & (Jy) & (Jy) & (Jy)\\ 
\hline                                          
G9.62+0.19     & 1  & $-20.8$ & $+10.3$ & \nodata & 0.20    & \nodata & \nodata \\             
               & 2  & $-14.5$ & $+11.8$ & \nodata & 1.56    & \nodata & \nodata \\
               & 3  & $-8.6 $ & $-7.8 $ & \nodata & 18.6    & \nodata & \nodata \\
               & 4  & $-6.8 $ & $+12.5$ & \nodata & 0.34    & \nodata & \nodata \\
               & 5  & $-5.2 $ & $+6.9 $ & \nodata & 0.14    & \nodata & \nodata \\              
               & 6  & $-2.6 $ & $+6.5 $ & \nodata & 0.17    & \nodata & \nodata \\               
               & 7  & $-1.8 $ & $+13.2$ & \nodata & 0.24    & \nodata & \nodata \\
               & 8\tablenotemark{\ddagger}  & $+1.6 $ & $-2.6 $ & 0.33\tablenotemark{c} & 0.43& 3.17\tablenotemark{c} & \nodata \\               
               & 9  & $+4.1 $ & $-12.3$ & \nodata & 0.07    & \nodata & \nodata \\
G11.94$-$0.62  & 1  & $+0.5 $ & $+0.5 $ & \nodata & 0.18    & \nodata & 0.83    \\
               & 2  & $+8.6 $ & $+0.8 $ & \nodata & 1.11\tablenotemark{?}   & \nodata & 5.43\tablenotemark{?}\\
               & 3  & $+10.8$ & $+5.0 $ & \nodata & 1.28\tablenotemark{?}   & \nodata & 7.82\tablenotemark{?}\\
               & 4  & $+11.1$ & $-2.3 $ & \nodata & 1.44\tablenotemark{?}   & \nodata & 11.3\tablenotemark{?}\\
               & 5  & $+11.7$ & $+0.1 $ & \nodata & 1.45\tablenotemark{?}   & \nodata & 11.1\tablenotemark{?}\\
               & 6  & $+17.3$ & $-3.3 $ & \nodata & 1.21\tablenotemark{?}   & \nodata & 8.70\tablenotemark{?}\\
               & 7  & $+17.7$ & $-7.2 $ & \nodata & $<$0.03 & \nodata & 5.01    \\
G19.61$-$0.23  & 1  & $-11.4$ & $+9.0 $ & $<$0.01 & 0.58    & 1.39    & 3.66    \\
               & 2  & $-8.0 $ & $+5.1 $ & 2.64    & 3.04    & 9.20    & 17.4    \\
               & 3  & $-3.5 $ & $-3.6 $ & 9.47    & 11.6    & 52.8    & 82.9    \\
               & 4  & $+5.6 $ & $+9.4 $ & 4.06    & 3.36\tablenotemark{?} & 18.3 & 33.5\tablenotemark{?}    \\
               & 5  & $+12.6$ & $-8.1 $ & 4.02    & 4.75\tablenotemark{?} & 21.2 & 20.9\tablenotemark{?}    \\
               & 6  & $+16.8$ & $-5.6 $ & $<$0.01 & $<$0.03 & 1.08    & 2.70    \\            
G34.26+0.15    & 1  & $-2.4 $ & $-1.2 $ & 0.24\tablenotemark{?g} & \em{U} & 0.64\tablenotemark{?g} & \em{U}  \\
               & 2  & $-2.1 $ & $-2.1 $ & 0.15\tablenotemark{?g} & \em{U} & 1.24\tablenotemark{?g} & \em{U}  \\
               & 3  & $-1.9 $ & $-0.4 $ & 0.51\tablenotemark{?g} & \em{U} & 0.75\tablenotemark{?g} & \em{U}  \\
               & 4  & $-1.3 $ & $+0.2 $ & 3.53\tablenotemark{?g} & \em{U} & 3.68\tablenotemark{?g} & \em{U}  \\
               & 5  & $+1.6 $ & $-1.4 $ & 0.81    & 0.61    & 2.06    & 5.11    \\
G45.07$-$0.13  & 1  & $-3.9 $ & $-7.3 $ & 0.89    & 1.40    & \nodata & 15.1    \\ 
               & 2  & $ 0.0 $ & $-2.2 $ & 20.4    & 21.7\tablenotemark{?} & \nodata & 19.7\tablenotemark{?}  \\
               & 3  & $+0.2 $ & $ 0.0 $ & 4.26    & 5.50\tablenotemark{?} & \nodata & 102\tablenotemark{?}  \\    
\hline
\end{tabular}
\vspace{-0.8cm}
\tablecomments{All 10.5 and 18.1 $\micron$ data were taken with OSCIR, and all 11.7 and 20.8 $\micron$ data were taken with MIRLIN. All data are from the IRTF unless otherwise noted. All values quoted with a ``$<$'' are upper limit flux densities at a 95\% confidence level. Flux density errors are taken to be 10\% at 10.5 and 11.7 $\micron$ and 15\% at 18.1 and 20.8 $\micron$ unless otherwise noted.}
\tablenotetext{\dagger}{Mid-infrared sources are numbered from west to east on each field. The positions of the sources are given as offsets in arcseconds from the maser/HMC location given in Table 1.}
\tablenotetext{\ddagger} {G9.619+0.193:DPT00 1 from De Buizer, Pi\~{n}a, \& Telesco (2000).}
\tablenotetext{c} {CTIO data from De Buizer, Pi\~{n}a, \& Telesco (2000).}
\tablenotetext{g} {Gemini data.} 
\tablenotetext{?} {There is some overlap of the extended emission from a nearby source. The sources are resolved enough that a value can be estimated, however we estimate the error in the flux densities to be 5\% higher in each filter.}
\tablenotetext{U} {The source was not sufficiently resolved from a nearby source enough to estimate a flux density measurement.}

\end{minipage}
\end{center}
\end{table*}   

\clearpage

\begin{figure}[t]
\figurenum{1}
\epsscale{0.50}
\plotone{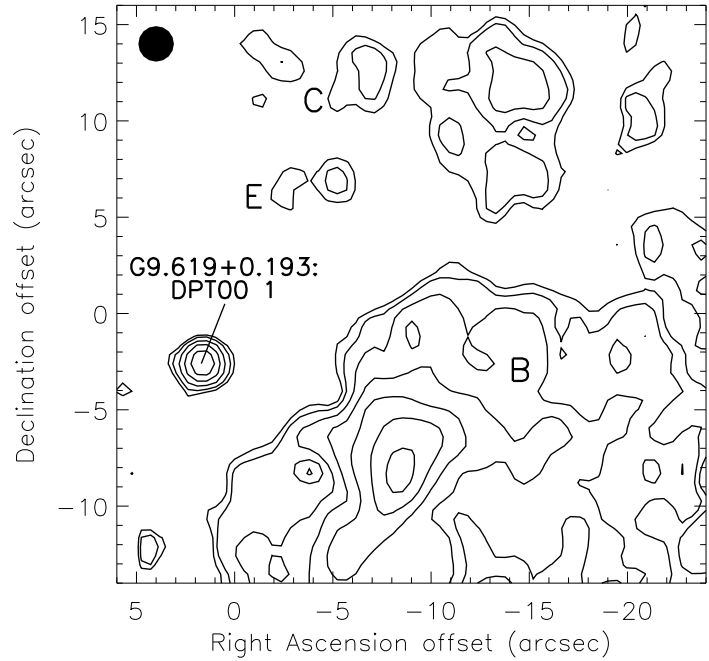}
\caption{Contour plot of the MIRLIN/IRTF 11.7 $\micron$ image of the G9.62+0.19 region. Labelled are the mid-infrared components radio sources as defined by Garay et al. (1993). Also labelled is the mid-infrared source G9.619+.193:DPT00 1. The filled circle in the upper left represents the FWHM resolution of the image.}
\end{figure} 

\begin{figure}[t]
\figurenum{2}
\epsscale{0.50}
\plotone{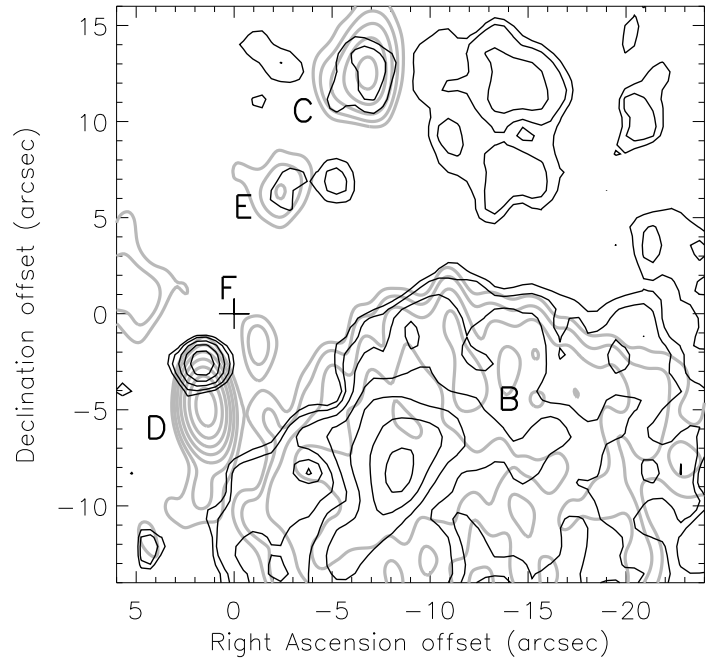}
\caption{A comparison between the 3.5 cm radio continuum image (gray contours) of Phillips et al. (1998) with the 11.7 $\micron$ IRTF image (black contours) taken with MIRLIN for the region around G9.62+0.19. Radio components as defined by Garay et al. (1993) are labelled. Sources C and E are imaged here for the first time in the mid-infrared. Along with the extended emission of source B, these sources finally and accurately anchor the astrometry between the radio continuum and mid-infrared. Radio source A is off-field and west of B. The position of the ammonia source center is shown as a cross and labelled F.}
\end{figure} 

\clearpage

\begin{figure}[t]
\figurenum{3}
\epsscale{0.50}
\plotone{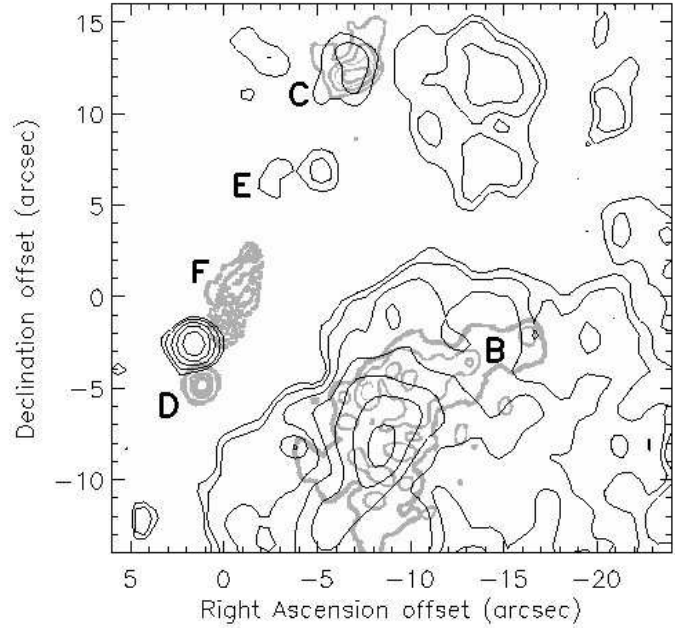}
\caption{A comparison between the 11.7 $\micron$ MIRLIN/IRTF image (black contours) and Figure 4 from Hofner et al. (1994) for G9.62+0.19. The gray continuous contours are high-resolution 3.6 cm radio continuum emission, and the gray broken contours delineate the areas of thermal NH$_3$(5,5) emission. In this image it can be clearly seen that the compact and bright mid-infrared source G9.619+0.193:DPT00 1 is not coincident with either the radio continuum source D nor the hot molecular core, F, seen in thermal ammonia.}
\end{figure}   

\begin{figure}[t]
\figurenum{4}
\epsscale{0.50}
\plotone{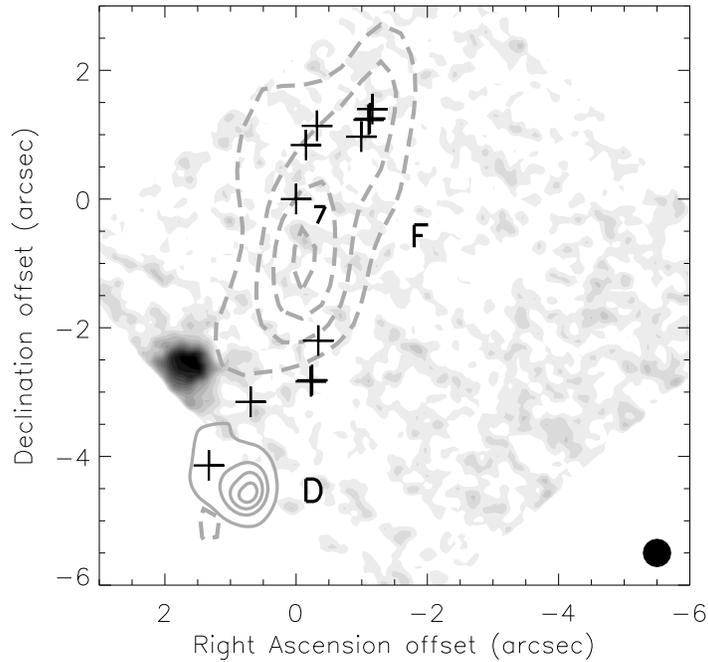}
\caption{The OSCIR/Keck 18.1 $\micron$ image (gray-scale) of G9.62+0.19 overlaid with the thermal ammonia emission (gray broken contours) of Hofner et al. (1994) of source F, and the 3.6 cm radio continuum emission (gray continuous contours) of source D from Testi et al. (2000). Also shown are the water maser group locations (crosses) from Hofner \& Churchwell (1996). Water maser group 7 is labelled and is the closest to the HMC center. No mid-infrared source was detected at the maser/HMC location with a 3-$\sigma$ upper limit value of 48 mJy at 18.1 $\micron$. The filled circle in the lower right represents the FWHM resolution of the mid-infrared Keck image.}
\end{figure}

\clearpage

\begin{figure}[t]
\figurenum{5}
\epsscale{0.50}
\plotone{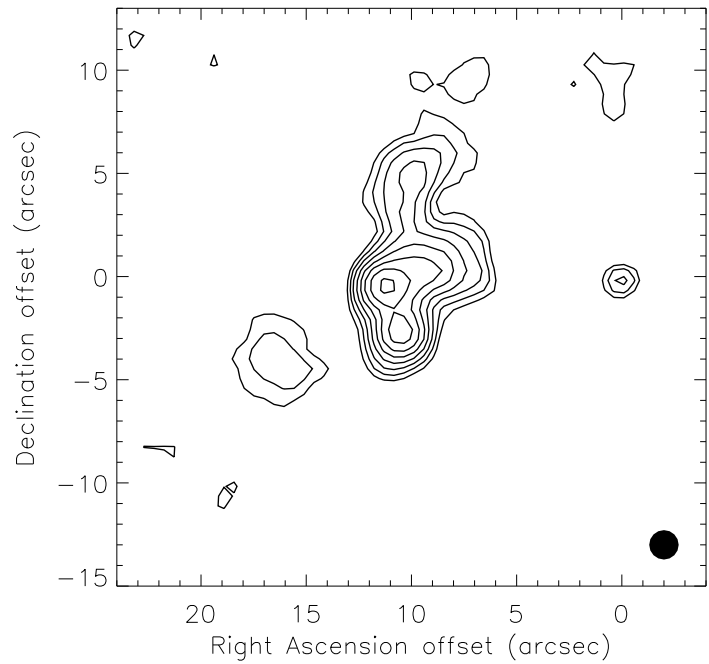}
\caption{A contour plot of the MIRLIN/IRTF 11.7 $\micron$ image of G11.94-0.62. The compact, unresolved source at the origin is the HMC candidate. The cometary UC H{\scriptsize II} region in this field does not appear to have a cometary shape in the mid-infrared. Instead it can be seen as the group of close and spatially unresolved sources to the east of the HMC location. The filled circle in the lower right represents the FWHM resolution of the image.}
\end{figure}

\begin{figure}[t]
\figurenum{6}
\epsscale{0.50}
\plotone{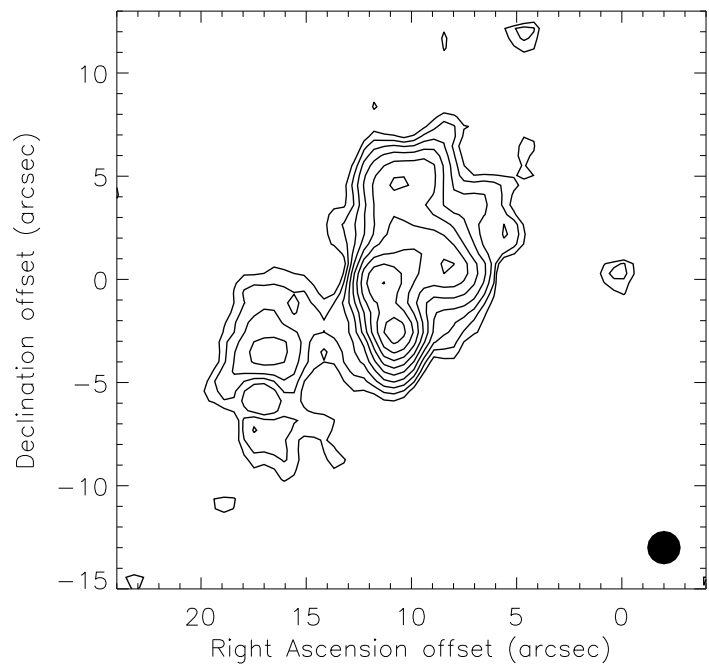}
\caption{A contour plot of the MIRLIN/IRTF 20.8 $\micron$ image of G11.94-0.62. Same comments as Figure 2a.}
\end{figure}

\clearpage

\begin{figure}[t]
\figurenum{7}
\epsscale{0.50}
\plotone{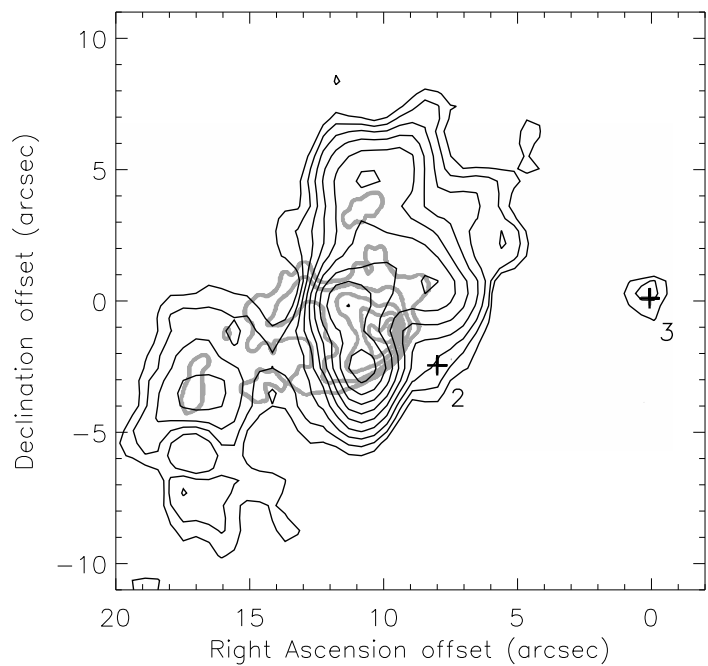}
\caption{A contour plot of the MIRLIN/IRTF 20.8 $\micron$ image of G11.94-0.62 (black contours) overlaid with the 2 cm radio continuum image (gray contours) of Hofner \& Churchwell (1996). Also shown are the numbered water maser group locations from Hofner \& Churchwell (1996). The peak of the mid-infrared emission from the HMC candidate is coincident with the location of maser group 3.}
\end{figure}

\clearpage

\begin{figure}[t]
\figurenum{8}
\epsscale{0.50}
\plotone{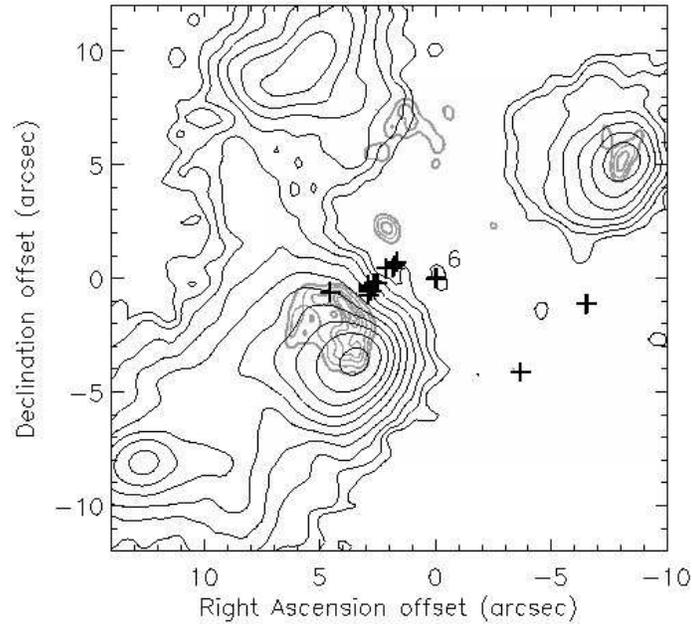}
\caption{A comparison between the 18.1 $\micron$ image (black contours) of G19.61-0.23 from De Buizer (2000) using OSCIR/IRTF and the 2 cm radio continuum image (gray contours) from Hofner \& Churchwell (1996). Also shown are the water maser locations from Hofner \& Churchwell (1996), with their water maser group 6 labelled. At this maser group location there is a 3-$\sigma$ detection of a mid-infrared source. Given the coincidence of the mid-infrared source with the water maser location and thermal ammonia source of Garay et al. (1998), this mid-infrared detection was considered to be real by De Buizer (2000).}
\end{figure}

\begin{figure}[t]
\figurenum{9}
\epsscale{0.50}
\plotone{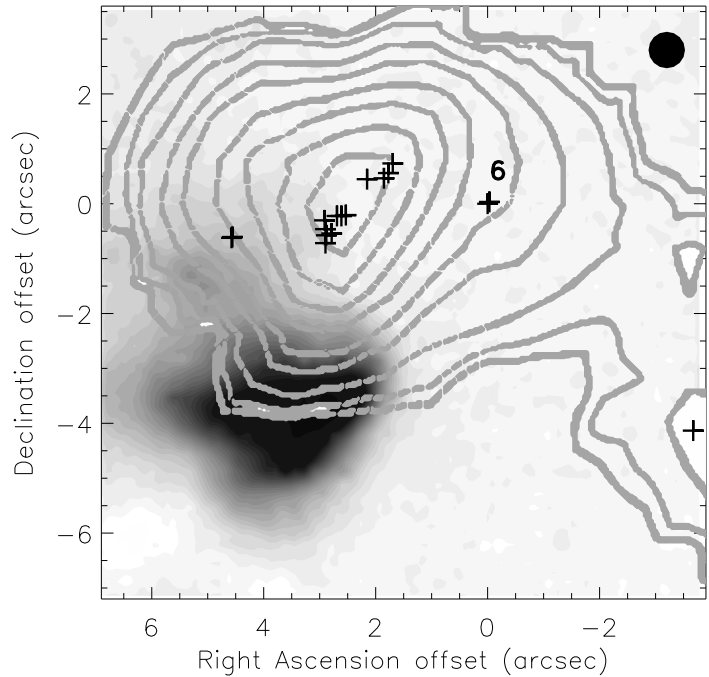}
\caption{Deeper 18.1 $\micron$ follow-up image of G19.61-0.23 from OSCIR/Gemini North (gray-scale). Overlaid in gray are the contours of the thermal ammonia source seen by Garay et al. (1998). The water masers of Hofner \& Churchwell (1996) are overlaid as well. There was no detection of the 18.1 $\micron$ source seen by De Buizer (2000) at the location of maser group 6. Therefore, it is likely that the HMC candidate seen previously was simply noise. The filled circle in the upper right represents the FWHM resolution of the mid-infrared Gemini image.}
\end{figure}

\clearpage

\begin{figure}[t]
\figurenum{10}
\epsscale{0.50}
\plotone{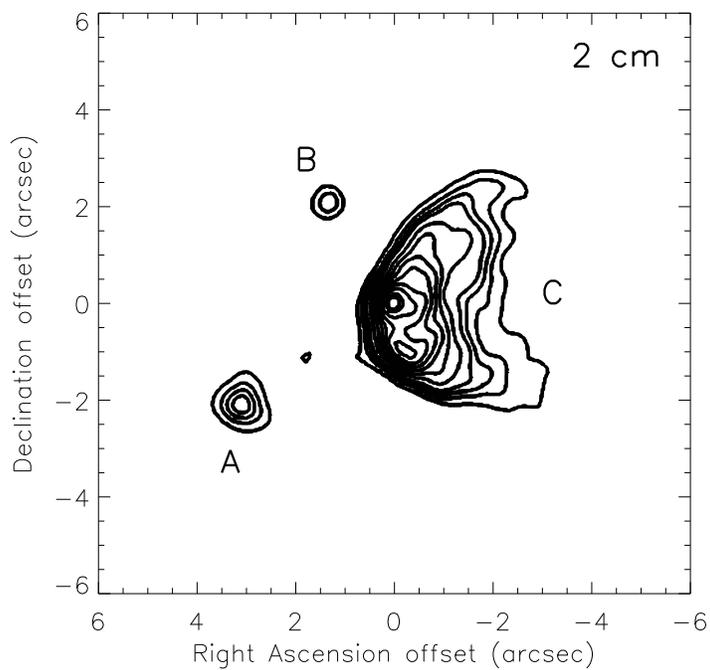}
\caption{A contour plot of the 2 cm radio continuum of G34.26+0.15 taken from Figure 5 of Hofner \& Churchwell (1996). This figure shows the radio continuum components in the region, which are labelled. The large and amorphous radio source D is not on field and is located $\sim$20$\arcsec$ to the southeast of A.}
\end{figure} 

\clearpage

\begin{figure}[t]
\figurenum{11}
\epsscale{0.50}
\plotone{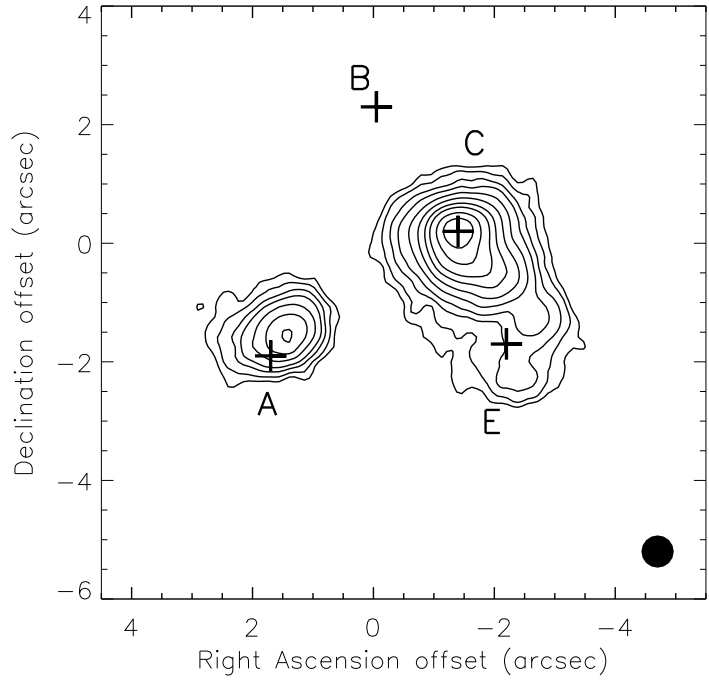}
\caption{A contour plot of the OSCIR/Gemini North 10.5 $\micron$ image of G34.26+0.15. Labelled crosses A-C mark the peaks of the radio continuum emission sources in the region. Previously detected mid-infrared source E from Campbell et al. (2000) is also marked with a cross. Like G11.94-0.62, the cometary UC H{\scriptsize II} region, C, in this field does not appear to have a cometary shape in the mid-infrared. It appears that source C and mid-infrared source E (Campbell et al. 2000) can be seen as four separate sources in this high resolution Gemini image. The filled circle in the lower right represents the FWHM resolution of the image.}
\end{figure}

\begin{figure}[t]
\figurenum{12}
\epsscale{0.50}
\plotone{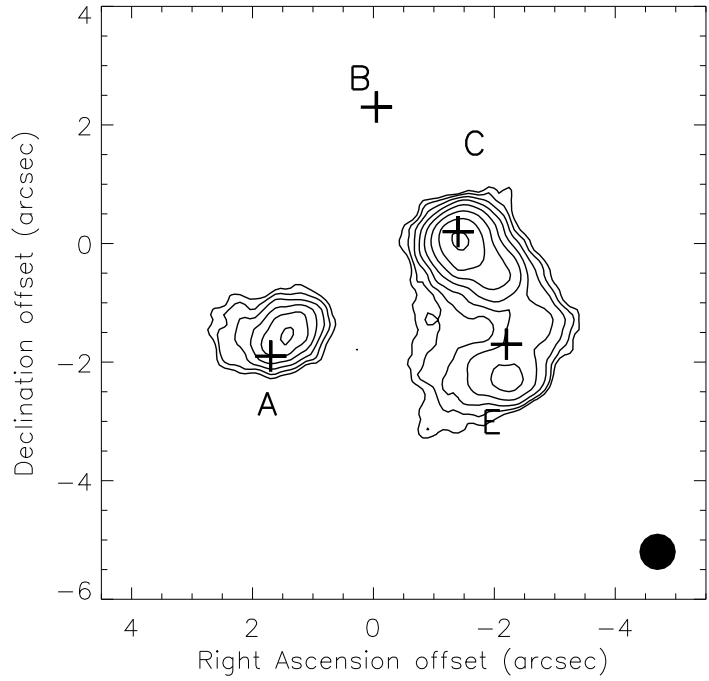}
\caption{The OSCIR/Gemini North 18.1 $\micron$ image of G34.26+0.15. Same comments as Figure 4a.}
\end{figure}

\clearpage

\begin{figure}[t]
\figurenum{13}
\epsscale{0.50}
\plotone{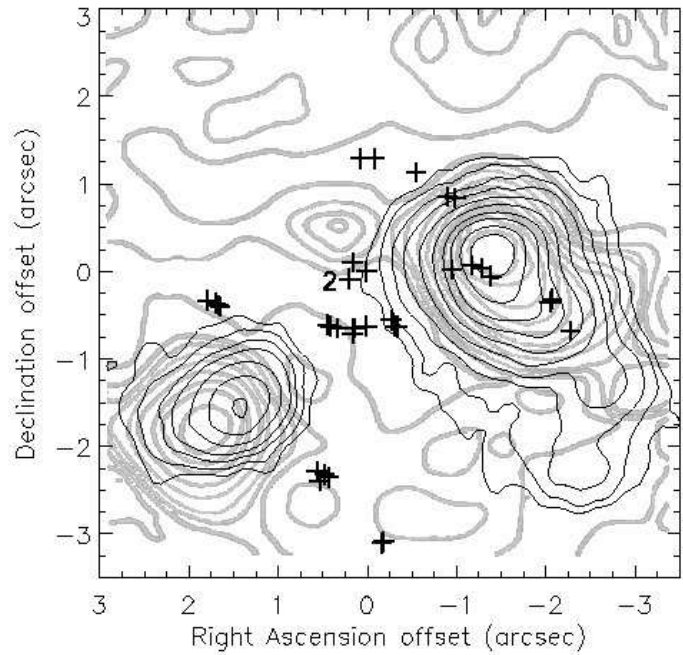}
\caption{The OSCIR/Gemini North 10.5 $\micron$ image of G34.26+0.15 (black contours) with the 12.5 $\micron$ image (gray contours) from Keto et al. (1992). Also shown are the water maser group locations from Hofner \& Churchwell (1996). Water maser group 2 is believed to be the site of an HMC. Keto et al. (1992) found a 70 mJy 12.5 $\micron$ source near the location of water maser group 2, claiming it to be mid-infrared emission from the HMC. We do not detect this source at 10.5 $\micron$ with a 3-$\sigma$ upper limit flux density of 4 mJy.}
\end{figure}

\begin{figure}[t]
\figurenum{14}
\epsscale{0.50}
\plotone{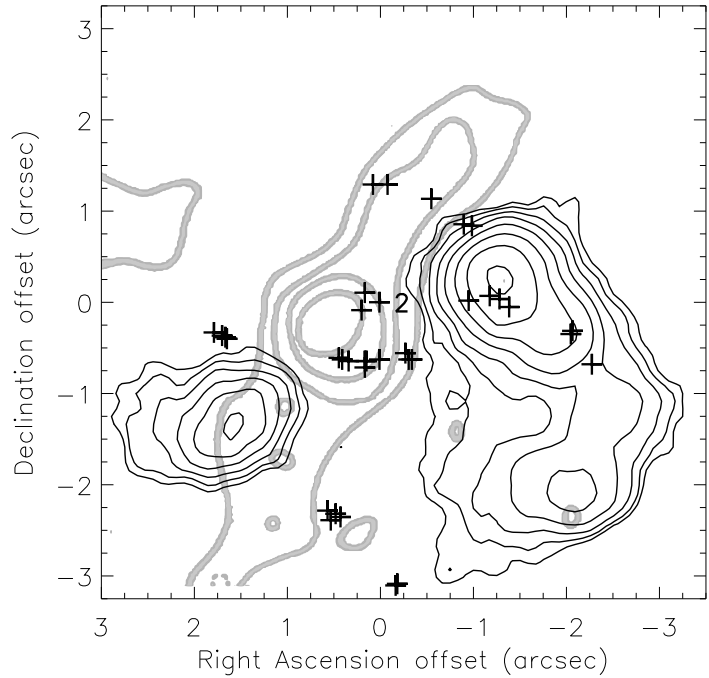}
\caption{The OSCIR/Gemini North 18.1 $\micron$ image of G34.26+0.15 (black contours) with the thermal NH$_3$ image (gray contours) from Keto et al. (1992). They find a bright thermal ammonia source near the location of water maser group 2. Keto et al. (1992) claim that the water maser group 2, the 12.5 $\micron$ source, and the ammonia source are spatially coincident, however they appear to be in three different locations. In any case, there is no 18.1 $\micron$ source at any of these three locations with a 3-$\sigma$ upper limit flux density of 21 mJy. }
\end{figure}

\clearpage

\begin{figure}[t]
\figurenum{15}
\epsscale{0.50}
\plotone{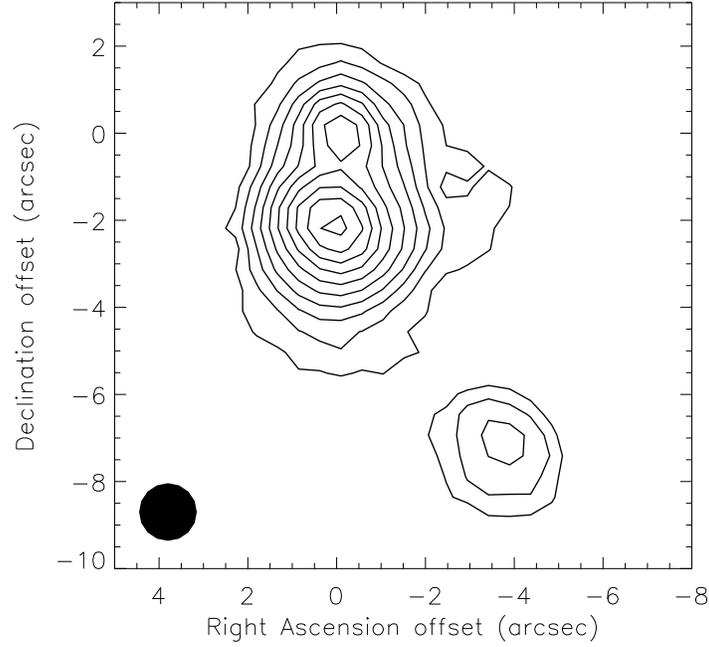}
\caption{A contour plot of the MIRLIN/IRTF 11.7 $\micron$ image of G45.07-0.13. Three mid-infrared sources were detected, including one at the candidate HMC location (origin). The filled circle in the lower left represents the FWHM resolution of the image.}
\end{figure}

\begin{figure}[t]
\figurenum{16}
\epsscale{0.50}
\plotone{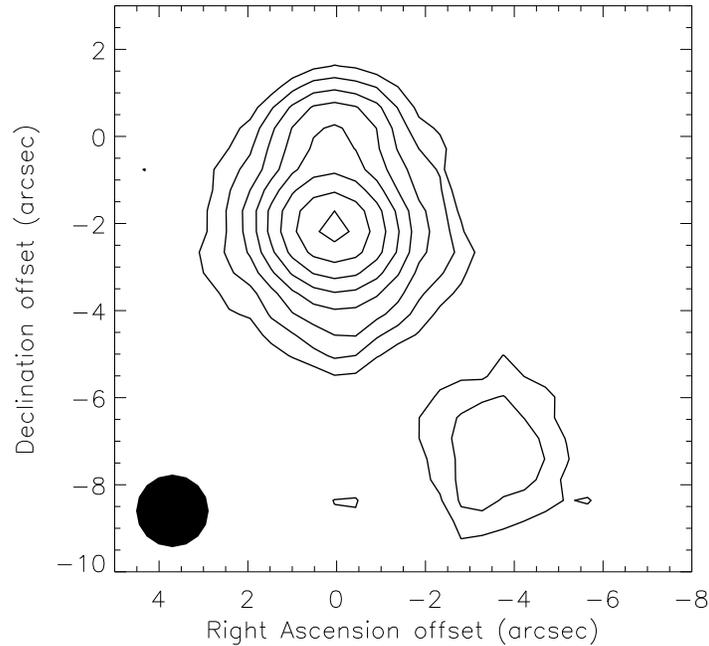}
\caption{A contour plot of the MIRLIN/IRTF 20.8 $\micron$ image of G45.07-0.13. Again three mid-infrared sources were detected, including one at the candidate HMC location (origin), however the two northern sources are not as spatially well-resolved as they are at 11.7 $\micron$. The filled circle in the lower left represents the FWHM resolution of the image.}
\end{figure}

\clearpage

\begin{figure}[t]
\figurenum{17}
\epsscale{0.50}
\plotone{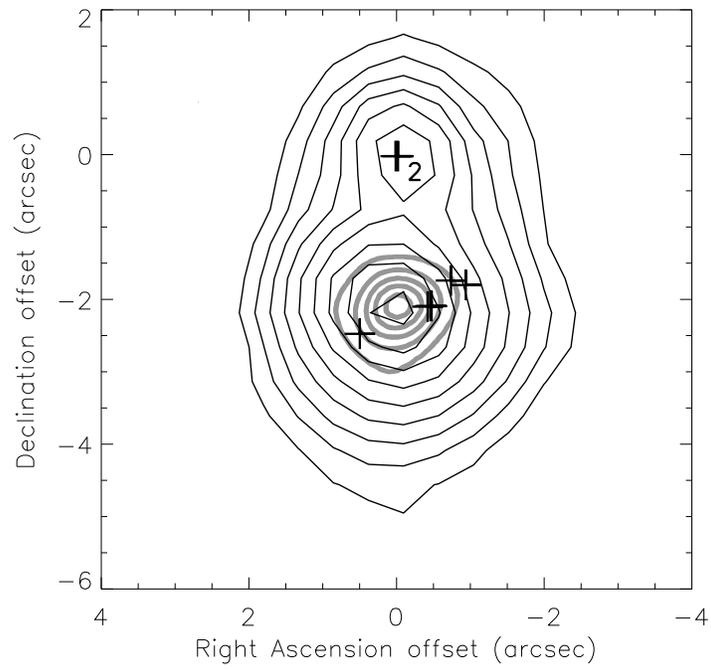}
\caption{The MIRLIN/IRTF 11.7 $\micron$ image of G45.07-0.13 (black contours) with the 2.0 cm radio continuum image of Hofner \& Churchwell (1996) overlaid (gray contours). Also shown are the water maser groups of Hofner \& Churchwell (1996). Maser group 2 is believed to mark the location of an HMC, and we have detected a bright mid-infrared source at this location. }
\end{figure}


\begin{thebibliography}{}

\bibitem[Anglada et al.(1996)]{1996ApJ...463..205A} Anglada, G., Estalella, 
R., Pastor, J., Rodriguez, L.~F., \& Haschick, A.~D.\ 1996, \apj, 463, 205

\bibitem[Campbell et al.(2000)]{2000ApJ...536..816C} Campbell, M.~F., 
Garland, C.~A., Deutsch, L.~K., Hora, J.~L., Fazio, G.~G., Dayal, A., \& 
Hoffmann, W.~F.\ 2000, \apj, 536, 816 

\bibitem[Carral et al.(1997)]{1997ApJ...486L.103C} Carral, P., Kurtz, 
S.~E., Rodriguez, L.~F., de Pree, C., \& Hofner, P.\ 1997, \apjl, 486, L103 

\bibitem[Cesaroni, Walmsley, \& Churchwell(1992)]{1992A&A...256..618C} 
Cesaroni, R., Walmsley, C.~M., \& Churchwell, E.\ 1992, \aap, 256, 618 

\bibitem[Cesaroni et al.(1994)]{1994A&A...288..903C} Cesaroni, R., 
Churchwell, E., Hofner, P., Walmsley, C.~M., \& Kurtz, S.\ 1994, \aap, 288, 
903 

\bibitem[Churchwell, Walmsley, \& Wood(1992)]{1992A&A...253..541C} 
Churchwell, E., Walmsley, C.~M., \& Wood, D.~O.~S.\ 1992, \aap, 253, 541 

\bibitem[De Buizer(2000)]{2000PhDT........D} De Buizer, J.~M.\ 2000, PhD 
Thesis, University of Florida

\bibitem[De Buizer(2003)]{2003gsfa.conf.....D} De Buizer, J.~M.\ 2003, in ASP Conference 
Series, Vol.~287, 
Galactic Star Formation Across the Stellar Mass Spectrum, J.~M.~De Buizer \& N.~S. van der Bliek, eds. (San Francisco: Astronomical Society of the Pacific), 230  

\bibitem[De Buizer, Pi{\~ n}a, \& Telesco(2000)]{2000ApJS..130..437D} De 
Buizer, J.~M., Pi{\~ n}a, R.~K., \& Telesco, C.~M.\ 2000, \apjs, 130, 437 

\bibitem[De Buizer et al.(2002a)]{2002ApJ...564L.101D} De Buizer, J.~M., 
Watson, A.~M., Radomski, J.~T., Pi{\~ n}a, R.~K., \& Telesco, C.~M.\ 2002, 
\apjl, 564, L101 

\bibitem[De Buizer et al.(2002b)]{2002ApJ...564L.101D} De Buizer, J.~M., 
Radomski, J.~T., Pi{\~ n}a, R.~K., \& Telesco, C.~M.\ 2002, 
\apj, 580, 305 

\bibitem[Garay, Rodriguez, Moran, \& Churchwell(1993)]{1993ApJ...418..368G} 
Garay, G., Rodriguez, L.~F., Moran, J.~M., \& Churchwell, E.\ 1993, \apj, 
418, 368 

\bibitem[Garay, Moran, Rodriguez, \& Reid(1998)]{1998ApJ...492..635G} 
Garay, G., Moran, J.~M., Rodriguez, L.~F., \& Reid, M.~J.\ 1998, \apj, 492, 
635 

\bibitem[Gaume, Fey, \& Claussen(1994)]{1994ApJ...432..648G} Gaume, R.~A., 
Fey, A.~L., \& Claussen, M.~J.\ 1994, \apj, 432, 648 

\bibitem[Hatchell, Thompson, Millar, \& 
MacDonald(1998)]{1998A&AS..133...29H} Hatchell, J., Thompson, M.~A., 
Millar, T.~J., \& MacDonald, G.~H.\ 1998, \aaps, 133, 29 

\bibitem[Hatchell et al.(2000)]{2000A&A...357..637H} Hatchell, J., Fuller, 
G.~A., Millar, T.~J., Thompson, M.~A., \& Macdonald, G.~H.\ 2000, \aap, 
357, 637

\bibitem[Hofner et al.(1994)]{1994ApJ...429L..85H} Hofner, P., Kurtz, S., 
Churchwell, E., Walmsley, C.~M., \& Cesaroni, R.\ 1994, \apjl, 429, L85 

\bibitem[Hofner \& Churchwell(1996)]{1996A&AS..120..283H} Hofner, P.~\& 
Churchwell, E.\ 1996, \aaps, 120, 283 

\bibitem[Hunter, Phillips, \& Menten(1997)]{1997ApJ...478..283H} Hunter, 
T.~R., Phillips, T.~G., \& Menten, K.~M.\ 1997, \apj, 478, 283 

\bibitem[Keto et al.(1992)]{1992ApJ...401L.113K} Keto, E., Proctor, D., 
Ball, R., Arens, J., \& Jernigan, G.\ 1992, \apjl, 401, L113 

\bibitem[Kruegel \& Walmsley(1984)]{1984A&A...130....5K} Kruegel, E.~\& 
Walmsley, C.~M.\ 1984, \aap, 130, 5 

\bibitem[Kurtz \& Franco(2002)]{2002RMxAC..12...16K} Kurtz, S.~\& Franco, 
J.\ 2002, in Revista Mexicana de Astronom{\'{\i}}a y Astrof{\'{\i}}sica Serie de 
Conferencias Vol.~12, Ionized Gaseous Nebulae, a Conference to Celebrate the 60th 
Birthdays of Silvia Torres-Peimbert and Manuel Peimbert, W.~J.~Henney, J.~Franco, M.~Martos, \& M.~Pe{\~ 
n}a), eds. (Morelia: UNAM),16

\bibitem[Millar \& Hatchell(1997)]{1997CoKon.100..207M} Millar, T.~J.~\& 
Hatchell, J.\ 1997, Communications of the Konkoly Observatory, 100, 207 

\bibitem[Olmi \& Cesaroni(1999)]{1999A&A...352..266O} Olmi, L.~\& Cesaroni, 
R.\ 1999, \aap, 352, 266 

\bibitem[Olmi, Cesaroni, \& Walmsley(1993)]{1993A&A...276..489O} Olmi, L., 
Cesaroni, R., \& Walmsley, C.~M.\ 1993, \aap, 276, 489 

\bibitem[Osorio, Lizano, \& D'Alessio(1999)]{1999ApJ...525..808O} Osorio, 
M., Lizano, S., \& D'Alessio, P.\ 1999, \apj, 525, 808 

\bibitem[Pankonin, Churchwell, Watson, \& 
Bieging(2001)]{2001ApJ...558..194P} Pankonin, V., Churchwell, E., Watson, 
C., \& Bieging, J.~H.\ 2001, \apj, 558, 194 

\bibitem[Persi et al.(2003)]{2003A&A...397..227P} Persi, P., Tapia, M., 
Roth, M., Marenzi, A.~R., Testi, L., \& Vanzi, L.\ 2003, \aap, 397, 227 

\bibitem[Phillips, Norris, Ellingsen, \& 
McCulloch(1998)]{1998MNRAS.300.1131P} Phillips, C.~J., Norris, R.~P., 
Ellingsen, S.~P., \& McCulloch, P.~M.\ 1998, \mnras, 300, 1131 

\bibitem[Plume, Jaffe, \& Evans(1992)]{1992ApJS...78..505P} Plume, R., 
Jaffe, D.~T., \& Evans, N.~J.\ 1992, \apjs, 78, 505 

\bibitem[Soifer et al.(2000)]{2000ApJ...119..509S} Soifer, B. T., Neugebauer, G., Matthews, K., Egami, E., Becklin, E. E., Weinberger, A. J., Ressler, M., Werner, M. W., Evans, A. S., Scoville, N. Z., Surace, J. A., Condon, J. J.\ 2000, \apj, 119, 509

\bibitem[Stecklum et al.(2002)]{2002A&A...392.1025S} Stecklum, B., Brandl, 
B., Henning, T., Pascucci, I., Hayward, T.~L., \& Wilson, J.~C.\ 2002, 
\aap, 392, 1025 

\bibitem[Testi, Felli, Persi, \& Roth(1998)]{1998A&A...329..233T} Testi, 
L., Felli, M., Persi, P., \& Roth, M.\ 1998, \aap, 329, 233 

\bibitem[Testi, Hofner, Kurtz, \& Rupen(2000)]{2000A&A...359L...5T} Testi, 
L., Hofner, P., Kurtz, S., \& Rupen, M.\ 2000, \aap, 359, L5 

\bibitem[Watt \& Mundy(1999)]{1999ApJS..125..143W} Watt, S.~\& Mundy, 
L.~G.\ 1999, \apjs, 125, 143 

\end{thebibliography}
\end{document}